\documentclass{article}

\usepackage[preprint]{neurips_2025}
\usepackage{packages}

\title{Humans expect rationality and cooperation from LLM opponents in strategic games}

\author{%
  Darija Barak\\
  School of Economics\\
  University of Edinburgh\\
  \texttt{dbarakh@ed.ac.uk} \\
  \And
    Miguel Costa-Gomes\\
    School of Economics\\
    University of Edinburgh\\
    \texttt{miguel.costa-gomes@ed.ac.uk}\\
}

\begin{document}

\maketitle

\begin{abstract}
As Large Language Models (LLMs) integrate into our social and economic interactions, we need to deepen our understanding of how humans respond to LLMs opponents in strategic settings. We present the results of the first controlled monetarily-incentivised laboratory experiment looking at differences in human behaviour in a multi-player \textit{p}-beauty contest against other humans and LLMs. We use a within-subject design in order to compare behaviour at the individual level. We show that, in this environment, human subjects choose significantly lower numbers when playing against LLMs than humans, which is mainly driven by the increased prevalence of `zero' Nash-equilibrium choices. This shift is mainly driven by subjects with high strategic reasoning ability. Subjects who play the zero Nash-equilibrium choice motivate their strategy by appealing to perceived LLM's reasoning ability and, unexpectedly, propensity towards cooperation. Our findings provide foundational insights into the multi-player human-LLM interaction in simultaneous choice games, uncover heterogeneities in both subjects' behaviour and beliefs about LLM's play when playing against them, and suggest important implications for mechanism design in mixed human-LLM systems.
\end{abstract}

\section{Introduction} \label{sec:introduction}
Large Language Models (LLMs) are swiftly moving beyond the role of mere chatbots, and are increasingly integrated into every dimension of our social and economic lives. Soon, they will inevitably become active participants of complex interactions which were previously reserved exclusively for humans. It is also not difficult to imagine that soon humans will be interacting in  environments dominated by hundreds and even thousands of agents with artificial intelligence (AI). This swift change presents a new critical research frontier: understanding how humans adjust their behaviour when faced with strategic environments populated not only by people but also AI agents. In a few years, understanding the nuances of human-AI strategic interaction will have profound implications for effective mechanism design in every facet of economic activity -- from markets for goods and services, through financial markets, to strategic negotiations.

To date, effort has been dedicated to understanding how LLMs behave in both standard economic games \cite{H_2023} and more complex environments \cite{PO_2023}. In contrast, strategic interactions of humans \textit{with LLMs} have been largely left unexplored, with the exception of the work by Engel et al. \cite{EGO_2024} which looks at how humans interact with LLMs in two-player dominance-solvable prisoner's dilemma. Crucially, an important question remains unaddressed: how do humans behave in strategic environments when faced with multiple LLM opponents where no dominant strategy exists? While existing experimental economics literature shows that human behaviour is sensitive to perceived characteristics of their opponents \cite{APST_2012, EKS_2008}, we currently lack insight into whether humans perceive LLMs as fundamentally different from humans -- more rational, or perhaps more predictable, or even cooperative, -- and how these perceptions feed into observable behaviour in complex strategic settings.

In this paper we take the first steps into the uncharted territory of human strategic interaction with multiple LLMs. To fix ideas, we focus on the classic \textit{p}-beauty contest -- a workhorse game for studying strategic reasoning (as revealed by subjects' action choices and beliefs), often seen as a simplified as a simplified model of financial markets \cite{N_1995,HCW_1998}. The core novelty of our approach is in using a within-subject experimental design, that allows us to compare how the same individuals approach the game when interacting with other human experimental participants or playing against LLMs (ChatGPT 3.5 and Claude v2). Our main interest here lies in looking at setups where humans interact with multiple opponents and no strategy for any of the players is dominant\footnote{A (weakly) dominant strategy is the strategy that is (weakly) better than any other strategy available to the player regardless of opponents' choices in the sense of expected payoff maximisation.}. Such setup implies that, to choose a strategy, our human participants have to form beliefs about their opponents' actions. This allows us to investigate whether humans perceive LLMs differently from other humans. 

We also utilise a version of the \textit{p}-beauty contests with $2$ contestants, which allows us to identify subjects with high strategic reasoning ability -- those that, beyond reasonable doubt, are capable of finding the (weakly) dominant strategy in the $2$-player game. Separation of subjects into subsamples based on their reasoning ability allows us to further study heterogeneities in human behaviour when dealing with LLMs as opposed to humans. We are also able to examine which beliefs drive these behavioural shifts. 

By carefully studying behaviour in a controlled laboratory environment with a design that is incentive-compatible, we provide foundational insights into human-LLM strategic interactions, revealing not only that humans treat LLMs differently, but beginning to uncover who does so, and why. Our results map direction understanding the dynamic of complex human-LLM strategic interactions and designing economic mechanisms of the future.

\section{Related work} \label{sec:related}
\textbf{Beauty contests with $\mathbf{2}$ and more players.} The behaviour of humans in $N=2$ and $N>2$ \textit{p}-beauty contests has been widely studied \cite{N_1995,GN_2008,HCW_1998,CMNP_2009}. However, no previous study has used a within-subject design to separate subjects based on their ability to `solve' the $2$-player version of the contest (by identifying the weakly dominant strategy) and examine behavioural heterogeneities between these subjects when facing different types of opponents. As explained in \Cref{sec:results}, in our experiment we are able to identify subjects with high strategic reasoning ability -- \textit{ie} those that can establish the (weakly) dominant strategy in the $2$-player game, -- and show that these subjects are disproportionally represented among those who adapt their behaviour when interacting with LLMs.

\textbf{Strategic behaviour and observable characteristics of opponents.} Experimental economics literature documents that subjects adjust their strategic behaviour depending on perceived characteristics of their opponents. For example, Agranov, et al. \cite{APST_2012} find that subjects choose relatively lower number in the \textit{p}-beauty contest when interacting with graduate students as opposed to undergraduate students. The central interest of this paper is in exploring how subjects interact with LLMs -- such as Chat-GPT, and Claude, -- which are rapidly becoming prevalent. To the best of our knowledge, no existing work has looked at human behaviour \textit{in the presence of LLMs} in such strategic environments where players face multiple opponents.

\textbf{Role of LLMs in researching human behaviour.} Leveraging LLMs to aid the study of human behaviour is a novel research direction. Most efforts focus on understanding how LLMs themselves behave in standard games of economic interest \cite{H_2023} or using LLMs to simulate human behaviour \cite{ABF_2023}. Research into the direct interaction of humans with LLMs is rarer, with the most prominent exception being the recent experiments on human-LLM interactions in the prisoners' dilemma \cite{EGO_2024}. None of the previous work has utilised a within-subject design to study differences in human strategic behaviour in settings with a single human and several LLM players, particularly in games like the \textit{p}-beauty contest with more than $2$ contestants which, unlike the prisoners' dilemma, lack a dominant strategy. The present paper is the first to address this gap.

\section{Theoretical framework} \label{sec:theory}
The \textit{p}-beauty contest is a standard tool in experimental economics for studying strategic reasoning and belief formation \cite{N_1995, HCW_1998}. The task amounts to choosing an integer between $0$ to $100$. The winner is the person whose choice is closest to the target number equal to a fraction $p$ of the average of all contestants' choices. Often, $p = 2/3$ but behaviour with other values of $p$, including those greater than one, has also been studied\footnote{Nagel \cite{N_1995} used values of $p$ between $0.5$ and $1.25$, while Ho et al. \cite{HCW_1998} studied values in the $0.7$-$1.3$ range.}. Crucially, the optimal guess in a \textit{p}-beauty contest depends on how many participants are there in the contest and the beliefs one forms about their opponents. 

\subsection{Identification of high strategic reasoning ability}

An important distinction between \textit{p}-beauty contests with $2$ or more than $2$ contestants that we use to identify subjects with high strategic reasoning ability lies in the existence of the (weakly) dominant strategy in the former, which is not present in the latter.

In \textbf{$\mathbf{2}$-player contests}, choosing zero is the unique weakly dominant strategy -- \textit{ie} the strategy that is (weakly) better than any other strategy against all possible strategies of one's opponent(s), -- for any $p < 1$ \cite{GN_2008}. Since the target number ($p \times \text{average}$) is always less than the average, the player choosing the lowest number, if unique, wins. Zero guarantees at least a tie against any opponent choice and a win against any choice larger than that. Identifying this requires recognising dominance of zero, either through iterative deletion of weakly dominated strategies or contingent reasoning. Despite this, many experimental subjects fail to choose zero \cite{GN_2008,CMNP_2009}, suggesting heterogeneity in strategic ability.

For \textit{p}-beauty contests with \textbf{more that two contestants}, no strategy is weakly dominant. While the target number cannot exceed $100p$ and so any strategies above $100p$ are weakly dominated, selecting among the undominated strategies depends entirely on a player's beliefs about their opponents' choices. Standard game theory predicts convergence to the Nash Equilibrium of zero through iterative deletion of weakly dominated strategies. The idea is that, if nobody is believed to be playing strategies above $100p$, then the maximum value of the target value becomes $100p^2$, which, for any $p < 1$ is smaller than $100p$. Such iterative reasoning is akin to the removal of weakly dominated strategies, which at the end leaves only one undominated strategy -- \textit{ie} zero\footnote{For $p = 2/3$, arriving to zero takes $11$ steps, while for $p = 1/2$ only $7$ steps are required.}. However, observed choices are typically far above zero \cite{N_1995, HCW_1998}, often interpreted through models of bounded rationality like Level-\textit{k} thinking \cite{N_1995,SW_1994,CGCB_2001,CHC_2004}. In this framework, players perform some, but limited, number of steps of iterative reasoning, with higher levels leading to lower choices.

This distinction between \textit{p}-beauty contests with $2$ or more than $2$ participants is key for our identification strategy in this study. The $2$-player version of the contest provides a benchmark for identifying subjects capable of recognising the (weak) dominance of the `zero' strategy, which is indicative of high strategic ability. On the other hand, the contest with more than $2$ contestants, lacking a dominant strategy, is well-suited for studying how beliefs about opponents' choices, and hence a player's own choices, could potentially be affected by the type of opponents (human or LLM) they face.

\subsection{The role of player's beliefs about their opponents' behaviour}

In \textit{p}-beauty contests with more than two contestants, as no strategy is (weakly) dominant, choosing a specific strategy requires forming beliefs about how others will play, and then best-responding to those beliefs in the sense of expected payoff maximisation. Standard models assume that players form beliefs about other humans, drawing on shared experiences or assumptions about human rationality and cognitive limits. For example, in a level-\textit{k} model it is conjectured that players of level-\textit{1} assume that their non-strategic opponents randomise uniformly on $[0,100]$ and best-respond to those beliefs, level-\textit{2} players best-respond to level-\textit{1} players \textit{etc}.

The introduction of LLM opponents fundamentally challenges standard strategic reasoning. While LLMs do not `think' in the usual sense of the word, the text generated by largest and most capable modern LLMs closely resembles human-generated text and thus may give an impression of a reasoning process. However, when dealing with LLMs, unlike when dealing with other humans, humans lack an established prior about LLMs behaviour in a given task. Should one expect LLMs to engage in unbounded reasoning and perform all steps of iterative elimination of (weakly) dominated strategies? Or is it more likely to mimic human play? Perhaps, one should expect LLMs to follow some other logic entirely, or be completely unpredictable? 

Understanding of human behaviour in human-LLM interactions requires examining both the behaviour itself but also the beliefs that humans form about their LLM opponents. Our experimental design described in \Cref{sec:design} allows us to examine how perceived opponent type (human or LLM) influences strategic choices and underlying reasoning, focusing on whether subjects of different strategic reasoning ability exhibit behavioural patterns when interacting with humans and LLMs.

\section{Experimental design} \label{sec:design}
The data for this paper was collected via a monetarily-incentivised economics experiment conducted in a laboratoryand is part of a larger experimental study which poses questions outside of the scope of this paper. Below, we describe parts of experimental design that are relevant to this paper. 

This study received ethical approval from the Ethics Review Board at the University. All subjects gave informed consent prior to participation and were free to withdraw from the study at any point. The design of the experiment was pre-registered\footnote{We used AsPredicted for pre-registration. A copy of the form is available upon request.}. In accordance with standard practices, the experiment was coded in o-Tree \citep{CSW_2016} with a server hosted on Heroku. 

\subsection{Elicitation of choices against other participants} \label{sec:design_humans}

The first part of the experiment consisted of a sequence of $6$ versions of the \textit{p}-beauty contest. The baseline variant -- \texttt{G3P2/3} -- was a $3$-player beauty contest with $p = 2/3$. The winner earned a prize of \pounds$3$, and in case of ties, the prize was split equally among the winners. For exact wording of the task, see \Cref{app:prompts}. For this paper, the other two relevant tasks were \texttt{G2P1/2}, \texttt{G3P1/2}, which are both variations of the above task with $p = 2/3$ or $1/2$ and group size set to $2$ players. All $6$ tasks (including the $3$ relevant for this paper) were presented in random order. 

\begin{figure}[t]
    \centering
    \frame{
        \includegraphics[width=\textwidth]{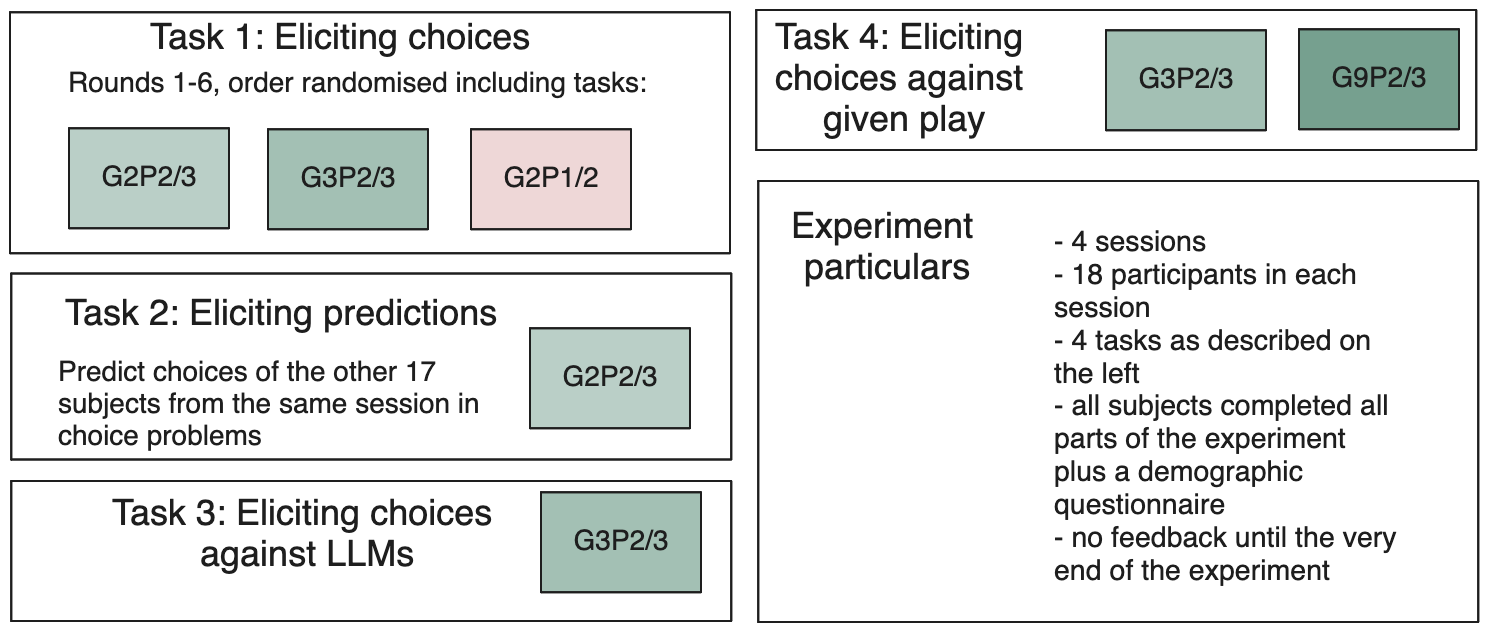}
    }
    \caption{Schematic summary of experimental design.}
    \label{fig:design}
\end{figure}

\subsection{Elicitation of predictions about choices of other participants} \label{sec:design_predictions}

As specified in \Cref{fig:design} (Task 2), in addition to eliciting subjects' choices in the various versions of the \textit{p}-beauty contest described above, in the second part of the experiment we elicited their predictions regarding the choices of other participants. For this paper, the relevant contest for which we elicited subjects' predictions is \texttt{G2P2/3}.  

Our predictions' elicitation task was similar to those proposed by Bosch-Rosa \& Meissner \cite{BRM_2020} and Lahav \cite{L_2015}, and amounted to asking subjects to allocate $17$ `tokens' -- one for every other subject participating in the same session -- across $20$ equal-size `bins'\footnote{Since p-beauty contest has $101$ possible choices, the first bin in our predictions elicitation task span numbers from $0$ to $5$ (inclusive). The second bin span numbers from $6$ to $10$, the third -- from $11$ to $15$, and so on.}. See \Cref{fig:predictions_interface} in \Cref{app:screenshots} for a screenshot of experimental interface for this task. 

We use the data from this task in \Cref{sec:reasoning_ability} to check whether subjects' actions are consistent with their stated beliefs about the actions of their opponents. That is, we check whether the choices subjects made in the \texttt{G2P2/3} version of the contest are best responses to their beliefs about choices of other experimental participants from the same session as elicited in this task.

\subsection{Elicitation of choices against LLMs} \label{sec:design_llms}

In the next task subjects were asked to make a choice in the \texttt{G3P2/3} version of the contest where their opponents were not other experimental participants but two Large Language Models (LLMs). The task presented to subjects in this part of the experiment was similar to that described in \Cref{sec:design_humans} but subjects were explicitly told that their two opponents are Chat-GPT (developed by OpenAI) and Claude (developed by Anthropic). For the exact wording of the task, see \Cref{app:prompts}.

To extract LLMs choices, we prompted both Chat-GPT (v3.5) and Claude (v2) with the same text. Both language models picked $50$ as their guess for the \texttt{G3P2/3} version of the $p$-beauty contest. The exact prompt used along with responses of both LLMs are in \Cref{app:llmsprompt}. This task was incentivised and we matched the choices of LLMs to choices of participants to determine earnings for this task. Notice that, when making their choices, our subjects had no knowledge of the actual choices of LLMs but were able to learn those at the very end of the experiment. 

\subsection{Elicitation of choices against given play} \label{sec:design_controls}

After the main part of the experiment, subjects completed two tasks where play of opponents was given to them. Subjects were asked to choose a decision in two versions of a $2/3$-beauty contest -- one with $G = 3$ and another one with $G = 9$. The decisions of their $2$ or $8$ opponents were drawn from $\mathcal{U}\{0, 100\}$ and given to the subjects, and the use of calculator was allowed. We included this task to check whether subjects are able to identify the best-response to the known choices of their opponents -- \textit{ie} an environment very similar to that in the experiment but with Knightian uncertainty and risk completely suppressed. We use the data from this task in \Cref{sec:reasoning_ability} to identify subjects who are able to best respond to known play of their opponents.

\subsection{Experimental workflow} \label{sec:design_workflow}

All sessions of the experiment unfolded in the same way as described below. Subjects read and signed the consent forms. Then they completed instructions and a quiz for the first task -- which amounted to making a sequence of decisions as described in \Cref{sec:design_humans} above. Next, subjects completed instructions and quiz for the second task, which was the predictions' elicitation task from \Cref{sec:design_predictions}. Next, as per \Cref{sec:design_llms}, subjects completed the task against LLMs, followed by the task described in \Cref{sec:design_controls}, and a short demographics survey. No feedback was provided until the very end of the experiment. After everyone completed all parts of the experiment, subjects learned their payoffs and were able to browse the history of their decisions and outcomes for all tasks. 

Subjects had access to an on-screen calculator throughout all parts of the experiment. In addition to standard operations, the calculator had buttons for multiplication by $1/2$ and $2/3$ to ease calculations if required. See \Cref{fig:calculator_interface} in \Cref{app:screenshots} for a screenshot of our on-screen calculator.

Subjects were paid for their choices in all parts of the experiment depending on their performance. Overall, subjects earned an average of \pounds$17$ inclusive of the \pounds$5$ show-up fee.

\subsection{Sample description} \label{sec:design_sample}

The experiment was conducted in April 2024 at the Behavioural Laboratory at the University using the local student population, including both undergraduates and postgraduates. A total of $72$ students completed the experiment across $4$ sessions. Of these, at the time of taking the experiment, $43$ students were studying towards an undergraduate degree, $22$ -- towards a postgraduate taught degree, and $3$ -- towards a PhD. The average age was $22.6$ years, and $54$ subjects were females. Each experimental session took about $1.5$ hours to complete.

When sampling, we avoided students from the Economics department, and focused on students from quantitative disciplines. Of the $72$ subjects, $1$ has taken at least one economics/game theory module as part of undergraduate education, and $43$ took at least one mathematics module. Finally, $5$ subjects indicated they came across the $p$-beauty contest or similar tasks before the experiment.

\section{Analysis and results} \label{sec:results}
In this section we analyse our data and present our results. First, in \Cref{sec:reasoning_ability}, we introduce conditions that subjects need to satisfy to be classified as displaying high strategic reasoning ability, and identify a subset of our sample who conform to these conditions. Second, in \Cref{sec:behavioural_differences}, we analyse the differences in behaviour of our subjects when interacting with other experimental participants and with LLMs. Here, we devote special attention to behavioural differences between subjects with high strategic reasoning ability and the rest of the sample. The insights from the qualitative data we collected on the motivations of behaviour against LLMs is discussed in \Cref{sec:llms_motivation}. All statistical tests reported below are two-tailed.

\subsection{High strategic reasoning ability} \label{sec:reasoning_ability}

In this section, using a systematic approach, we identify a subset of subjects characterised by high strategic reasoning ability. To achieve this, we specify and justify three conditions that a subject needs to fulfil in order to be classified as possessing high strategic reasoning ability. 

We classify subjects as possessing high strategic reasoning ability if they simultaneously satisfy three conditions designed to identify genuine understanding of the game structure beyond chance:
\begin{itemize}[nosep]
    \item[A:] Chose the weakly dominant strategy (zero) in at least one 2-player contest -- \textit{ie} \texttt{G2P2/3} and \texttt{G2P1/2} (\Cref{sec:design}, Task \#1).
    \item[B:] Correctly best-responded to induced -- \textit{ie} known and given, -- opponents' choices (\Cref{sec:design}, Task \#5).
    \item[C:] Exhibited choices consistent with their stated predictions about choices of other experimental participants in the \texttt{G2P2/3} task  (\Cref{sec:design}, Task \#1 and Task \#2). 
\end{itemize}

Compliance with all three conditions suggests a robust understanding unlikely to occur by chance. For context, $6$ of our subjects ($8.3\%$) chose the dominant strategy in both $2$-player contests -- an outcome with extremely low probability under uniform randomisation ($p < (1/101)^2$), -- and all of these individuals also met conditions B and C, thus falling within our high strategic reasoning ability classification. For full details of our classification method, see \Cref{app:reasoning_detail}.

Overall, we identify that $16.7\%$ of subjects in our sample can be classified as possessing high strategic reasoning ability. Importantly, all of them play zero in at least one of the $2$-player versions. 

\begin{Resbox}{Result 1}
    In our sample, $16.7\%$ of subjects possess high strategic reasoning ability. Crucially, these are the subjects who are capable of `solving' the $2$-payer version of the \textit{p}-beauty contest by identifying the (weakly) dominant strategy. 
\end{Resbox}

\subsection{Differences in behaviour against other subjects and LLMs} \label{sec:behavioural_differences}

\begin{figure}[t]
    \centering
    \frame{
        \includegraphics[width=\textwidth]{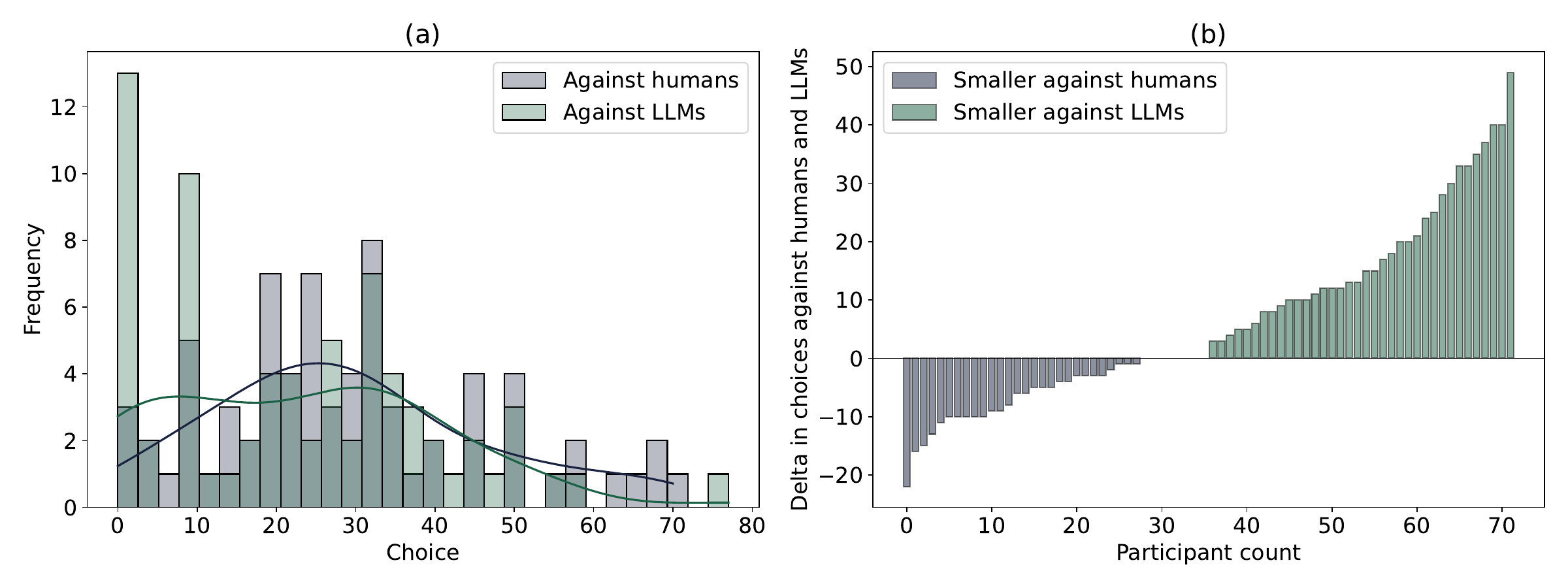}
    }
    \caption{\texttt{G3P2/3} task when playing against humans and LLMs: \textbf{(a)} Frequency histograms of observed choices and \textbf{(b)} subject-specific differences between choices.}
    \label{fig:humans_llms}
\end{figure}

\Cref{fig:humans_llms}(a) presents histograms of subjects' choices in the `humans' and LLMs versions of the \texttt{G3P2/3} task, both with fitted kernel density estimates. The mean number that subjects chose in the human version of the task was $29$ (std $17.3$). When interacting with LLMs, the mean number was $23$ (std $17.3$). This difference is statistically significant \textit{(Wilcoxon Sign Rank test (WSR), p-val $=$ 0.004, N$=$72)}. 

\Cref{fig:humans_llms}(b) presents subject-specific differences between choices submitted against humans and against LLMs. We see that, choices of $50\%$ of our subjects against LLMs were lower than their choices against humans. For those subjects, the average reduction in the chosen number is $18.2pts$. In addition, $12.3\%$ of subjects submitted the same choices against both humans and LLMs. Finally, for the remainder of the sample, the average increase in the chosen number is $7.3pts$.

\begin{Resbox}{Result 2}
    At the individual level, choices of subjects in \textit{p}-beauty contests made against LLMs are statistically lower than those made against humans. Importantly, we observe that $50\%$ -- \textit{ie} the plurality, -- of our subjects submitted lower choices against LLMs than against humans, while considerably fewer -- \textit{ie} $37.7\%$, -- submitted relatively higher choices against LLMs. Te average decrease in the chosen number in the former subsample was almost $2.5$ times larger in magnitude than the average increase in the latter.
\end{Resbox}

Next, from \Cref{fig:humans_llms}(a), we see that there are significantly more zero-choices made against LLMs than against humans. The share of subjects we observe choosing zero when interacting with real people in the \texttt{G3P2/3} version of the $p$-beauty contest is $4.2\%$, rising to $15.3\%$ in the LLMs version. This difference is statistically significant \textit{(WSR, p-val $=$ 0.024, N$=$72)}. Interestingly, of these $15.3\%$ zero-decisions, $36.4\%$ come from subjects who display high strategic reasoning ability who themselves represent only $16.7\%$ of the sample.

In fact, most of the difference between the choices against humans and LLMs is driven by the frequency of zero choices. In particular, discarding zeroes, the average choice we observe in the human version of \texttt{G3P2/3} task is $30.5$, and $27.2$ -- in the LLMs version of the task. This is a considerable reduction from the `raw' difference of $6pts$ observed when all observations are included. Nevertheless, dropping $20$ observations which have at least one choice of zero across the two tasks, the difference between the choices becomes only \textit{marginally} significant \textit{(WSR, p-val $=$ 0.060, N$=$52)}. 

We recognise that the above results may be `muddled' by introspective learning. After all, all human-to-human interactions in the experiment came strictly before any interactions with LLMs. However, in $3$ out of $4$ of our sessions, the relevant choice against humans in the \texttt{G3P2/3} task was made in round $4$ or $5$ -- \textit{ie} relatively late in the sequence of decisions once introspective learning has plausibly already happened. Only in $1$ of our sessions did this decision come first in the sequence. 

Recall that the difference in mean choices in the human and LLM  versions of the task is $6pts$. Excluding data from $18$ subjects from the session where the \texttt{G3P2/3} task came first, the difference between mean choices of subjects in the two versions of the tasks falls to $4.1pts$ and becomes insignificant with a two-tailed test \textit{(WSR, p-val $=$ 0.153, N$=$54)}. However, we see that in the $3$ relevant sessions, zero is selected in $16.7\%$ of decisions made against LLMs relative to $5.6\%$ of decisions made against humans \textit{(WSR, p-val $=$ 0.066, N$=$54)}. Overall, our data suggests that subjects choose zero significantly more often when they interact with LLMs as opposed to humans.

The last question we ask is whether the behaviour of the subjects with high strategic reasoning ability is different from the rest of the sample when it comes to interaction with LLMs. Observe that the $3$-player version of the \textit{p}-beauty contest does not have a weakly dominant action, so it is not immediately obvious that we should see subjects with high strategic reasoning ability choosing zero relatively more often. Yet, this is exactly what we see with $35.7\%$ of these subjects playing zero against LLMs relative to $10\%$ observed in the rest of the sample \textit{(Mann-Whitney U-test, p-val $=$ 0.066, N$_1=$17, N$_2=$55)}.

\begin{Resbox}{Result 3}
    The previous result is mostly driven by subjects choosing zero more often when playing the \textit{p}-beauty contest against LLMs compared to the case where they interact with other humans. The observation persists when introspective learning is accounted for. Subjects with high strategic reasoning ability are more likely to play zero against LLMs relative to the rest of the sample.
\end{Resbox}

This result suggests that, at least in one-shot interactions with LLMs, people -- especially those with high strategic reasoning ability, -- have a tendency to perceive these models differently compared to other experimental participants and appear to expect more equilibrium behaviour from LLMs.

\subsection{Motivations of behaviour against LLMs} \label{sec:llms_motivation}

Once subjects made their decisions against LLMs, we asked them to explain their reasoning in this part of the experiment. This was an open-ended non-incentivised question.  Focusing on $11$ subjects who chose zero in the LLM task, $6$ respond by referring to LLMs ability to `reason' through to select a zero -- \textit{eg} through familiarity with similar tasks, explicit iterative reasoning, \textit{etc}. Interestingly, other $4$ subjects refer to cooperation in their explanations. In particular, they say that they expect LLMs to select zero as, in the case of successful cooperation, this would guarantee that multiple winners share the prize equally\footnote{Observe that all players choosing the same number -- whether it is 0, 1, or 25, -- guarantees that the prize is split among them equally. The fact that players who expect cooperation from LLMs choose zero rather than any other number means they also recognise that this is an equilibrium outcome where nobody has an incentive to deviate.}. One subject explicitly states that they expect LLMs to cooperate at least with each other. 

\begin{Resbox}{Result 4}
    When interacting with LLMs in \textit{p}-beauty contests with no weakly-dominant strategy, subjects who choose zero do so because they expect LLMs to also choose zero. While most of these subjects expect LLMs to be able to reason through to equilibrium, another $36\%$ expect LLMs to display cooperative behaviour.
\end{Resbox}

This suggests that people may perceive LLMs not only as being more prone to behave in a manner consistent with Nash equilibrium but also perhaps display more cooperation relative to humans. The perception of LLMs as prone to cooperation is particularly noteworthy in the context of the \textit{p}-beauty contest -- an inherently non-cooperative setting where coordinated action does not lead to increased individual or collective benefit. 

It is worth pointing that out of the $6$ subjects who appeal to LLMs ability to reason, $3$ belong to our subset of subjects with high reasoning ability, whereas out of the $4$ subjects who quote cooperative tendencies, $3$ belong to this subset. Intriguingly, these subjects, who are themselves highly able strategically, appear capable of attributing both high-level strategic reasoning and seemingly non-strategic cooperative tendencies to LLMs, suggesting non-trivial mental models of LLM opponents.

\section{Discussion} \label{sec:discussion}
In this section, we discuss several questions which arose from the experiment we report in this paper. 

\textbf{Who are those who treat LLMs differently to humans in strategic settings?} While we observe $50\%$ of our sample choosing lower numbers when interacting with LLMs rather than humans, we find that subjects with high strategic reasoning ability are disproportionally represented among those treating LLMs differently to human opponents in \textit{p}-beauty contests. These subjects, capable of `solving' the $2$-player version of the contest, were more likely to choose zero when facing LLMs compared to the rest of the sample. This suggests that, whether humans treat LLMs differently to other humans in strategic settings may significantly interact with their own strategic sophistication. While our findings point to strategic sophistication as the main factor driving behavioural heterogeneities, these results stem from a university sample drawn primarily from quantitative backgrounds. Further research is needed to understand how our findings generalise to the broader population.

\textbf{What are the drivers of those differences?} The main driver appears to be the expectation that LLMs are more likely to possess greater strategic sophistication, thereby displaying equilibrium-like behaviour. Using qualitative data, we find that many subjects who play zero anticipate LLMs to be able to reason through to zero. Importantly, a notable proportion also cited perceived tendency to cooperation as a reason, expecting models to aim to share the prize equally. This is particularly suprising in a \textit{p}-beauty contest setting which is not a cooperative setting. We also find that subjects with high strategic reasoning ability are disproportionally represented among those citing both reasoning and cooperation motivations. Notice however that these beliefs were formed based on one-shot \textit{p}-beauty contests with specific LLMs (ChatGPT 3.5, Claude v2). How beliefs form in other strategic settings and repeated interactions with feedback are both open questions.

\textbf{What are the implications for mechanism design?} Our results highlight that, when designing mechanisms for mixed human-LLM interactive settings, we must account for heterogeneity of various types of agents as well as their beliefs about opponents. Assumptions about human behaviour using data from strategic environments populated only by humans may fail in settings where (numerous) LLM agents are involved. Researchers working on mechanism design need to carefully investigate the validity of those assumptions and ensure the mechanisms of the future are robust to various beliefs about LLM `intentions' and reasoning capabilities. At the same time, it may be possible to leverage the perception of LLM sophistication or cooperativeness to guide human behaviour. While the potential implications for mechanism design illuminated in this paper arise from observations of human behaviour specifically within \textit{p}-beauty contests, they highlight the broader challenge of designing mechanisms for human-AI settings. Further research is needed to establish whether our findings translate into other environments such as (anti-)coordination or bargaining games.

\textbf{What are the implications for future research?}  Future research on this topic should significantly broaden the scope. Effort should be directed at exploring human-LLM interactions in a wider range of strategic environments -- such as (anti-)coordination, public goods, and bargaining games, -- and in dynamic settings with learning opportunities. Importantly, research should test interactions with diverse LLMs beyond the two versions we used, including newer architectures and models calibrated for strategic play. Looking at the effects of scaling the number of LLM agents, potentially to hundreds or even thousands, and understanding how sensitive human behaviour is to varying LLM prompting strategies would also be valuable. 

\section{Conclusion} \label{sec:remarks}
This paper is the first to report results of a controlled experiment where human subjects interact with Large Language Models (LLMs) in strategic environments. We show that, in \textit{p}-beauty contests with no (weakly) dominant strategy, more subjects choose to play zero against LLMs than against other human participants. This behaviour suggests that subjects expect greater strategic sophistication from LLMs. Indeed, when explaining motivation for their zero-choices against LLMs, most subjects said that they expect models to be able to reason through to zero. Interestingly, the second motivation for choosing zero that we identify is the expectation that LLMs are more cooperative. Importantly, subjects who possess high strategic reasoning ability are disproportionally represented among those who chose to play zero against LLMs. While some research effort have already been directed to understanding how LLMs behave in both single-player and strategic environments, our paper is the first to illuminate the unexplored area of human interaction with LLMs in strategic settings with no (weakly) dominant strategy. We argue that understanding how and why humans interact with LLMs in strategic settings is an important emerging research direction.

% \section*{Acknowledgement}

% \begin{ack}
% Use unnumbered first level headings for the acknowledgments. All acknowledgments

% Do {\bf not} include this section in the anonymized submission, only in the final paper. You can use the \texttt{ack} environment provided in the style file to automatically hide this section in the anonymized submission.
% \end{ack}

% \section*{References}

% References follow the acknowledgments in the camera-ready paper. Use unnumbered first-level heading for the references. Any choice of citation style is acceptable as long as you are consistent. It is permissible to reduce the font size to \verb+small+ (9 point) when listing the references. Note that the Reference section does not count towards the page limit.
\medskip

{
\small

\bibliographystyle{abbrv}
\bibliography{bibl}

%%%%%%%%%%%%%%%%%%%%%%%%%%%%%%%%%%%%%%%%%%%%%%%%%%%%%%%%%%%%

\appendix

\section{Wording of human and LLM versions of the contest} \label{app:prompts}
The wording of the human version of the \texttt{G2P2/3} task was as follows:  

\textit{Choose a number between 0 and 100 (inclusive of both endpoints). You will be matched in a group with 2 other anonymous participants from this session. The winner will be the person whose number is closest to the target number. The target number is calculated as $2/3$rds of the average of all chosen numbers of the 3 people in your group (including you). The winner will get a prize of \pounds$3$. In case of a tie the prize will be split equally among those who tie.}

In contrast, below is the wording of the LLM version of the \texttt{G2P2/3} task as presented to our human subjects:

\textit{Choose a number between 0 and 100 (inclusive of both endpoints). You will be matched in a group with 2 other anonymous participants from this session. The winner will be the person whose number is closest to the target number. The target number is calculated as $2/3$rds of the average of all chosen numbers of the 3 people in your group (including you). The winner will get a prize of \pounds$3$. In case of a tie the prize will be split equally among those who tie.}

\textit{Further suppose that the 2 anonymous players you are interacting with are advanced language models. In particular, you are playing against two advanced generative artificial intelligence chatbots: Chat-GPT (developed by OpenAI) and Claude (developed by Anthropic).}

\section{Prompt for LLMs and models' responses} \label{app:llmsprompt}
We prompted Claude (v2, Anthropic) and Chat-GPT (v3.5, OpenAI) on 07/02/2024 with the following text: 

\textit{Suppose you face the following decision scenario.  You need to choose a number between 0 and 100 (inclusive of both endpoints). You will be matched in a group with 2 other anonymous players. The winner will be the player whose number is closest to the target number. The target number is 2/3rds of the average of all chosen numbers of the 3 players in your group (including you). The winner will get a prize of $\pounds$3. In case of a tie the prize will be split equally among those who tie. Choose your number.}

Claude chose $50$ and responded as follows: ``Okay, let's think through this strategically. The target number will be 2/3 of the average of the numbers chosen by me and the other 2 players. If I choose 100, the maximum possible number, the average could be at most (100 + 100 + 100)/3 = 100. So 2/3 of that is about 67. However, if the other players choose lower numbers, the average will be lower and so will the target number. So choosing 100 runs the risk that my number will be much higher than the target. On the other hand, if I choose 0, the minimum possible number, the average could be at most (0 + 100 + 100)/3 = 67. So 2/3 of that is about 44. But again, if the other players choose higher numbers, my number of 0 may end up too low compared to the target. So I think the optimal strategy is to choose 50. This balances the risk of being too high or too low compared to the range of possible target numbers. 50 seems like a reasonable guess at what 2/3 of the average might be. Therefore, I choose 50 as my number."

Chat-GPT also chose $50$ and provided the following response: ``In this scenario, a rational approach involves considering what other players might choose and trying to optimize your own chances of winning or at least tying for the prize. Since the target number is 2/3rds of the average of all chosen numbers, we can expect players to choose numbers that are higher than 0 and lower than 100, as choosing extreme values wouldn't be advantageous. One common strategy in such scenarios, known as the "centipede game," involves reasoning that players should aim to choose a number around the middle of the possible range, as this would likely lead to a higher average and thus a higher target number. Given this, a reasonable choice might be around 50, as it's in the middle of the range and likely to be close to the average of the numbers chosen by all three players. So, I would choose 50."

\section{Identification of subjects with high strategic reasoning ability} \label{app:reasoning_detail}
In this appendix we specify and explain the three conditions we use to identify subjects with high strategic reasoning ability. To be classified as someone with such ability, a subject must simultaneously satisfy all three conditions. \Cref{fig:high_reasoning} schematically shows the subsets of our sample which conform with various combinations of conditions A, B, and C.

\textbf{Condition A: choose weakly dominant strategy at least once.} Our first requirement is that a subject with high strategic reasoning ability should choose the weakly dominant strategy -- \textit{ie} play zero, -- in at least one contest with $2$ players. The rationale is that people may not `get it right' the first time they face such a contest, but arrive to this strategy through introspective learning. In our sample, in addition to the $6$ ($8.3\%$) subjects who choose zero in both $2$-player contests, $10$ subjects ($13.9\%$) chose zero in one of them only\footnote{We also observe that only one of the subjects who chose zero in the first $2$-player version of the contest they faced in the random sequence of contests subsequently chose a \textit{different number} in the second $2$-player version. That is, only one subject, for some reason, has `unlearned' the weakly dominant strategy through introspection. This subjects failed to satisfy one other criteria we impose -- \textit{ie} best responding to stated beliefs in \texttt{G2P2/3}, -- and so is not classified as possessing high strategic reasoning ability.}. 

Thus, in addition to the $6$ subjects who chose zero in both \texttt{G2P1/2} and \texttt{G2P2/3}, we have $10$ of subjects who may have `solved' the contest at their second attempt thanks to introspective learning. However, it may also be that they chose zero the second time purely `by accident'. To rule out accidental choices, we make impose two extra conditions as follows.

\begin{figure}[t]
    \centering
    \frame{
        \includegraphics[width=0.8\textwidth]{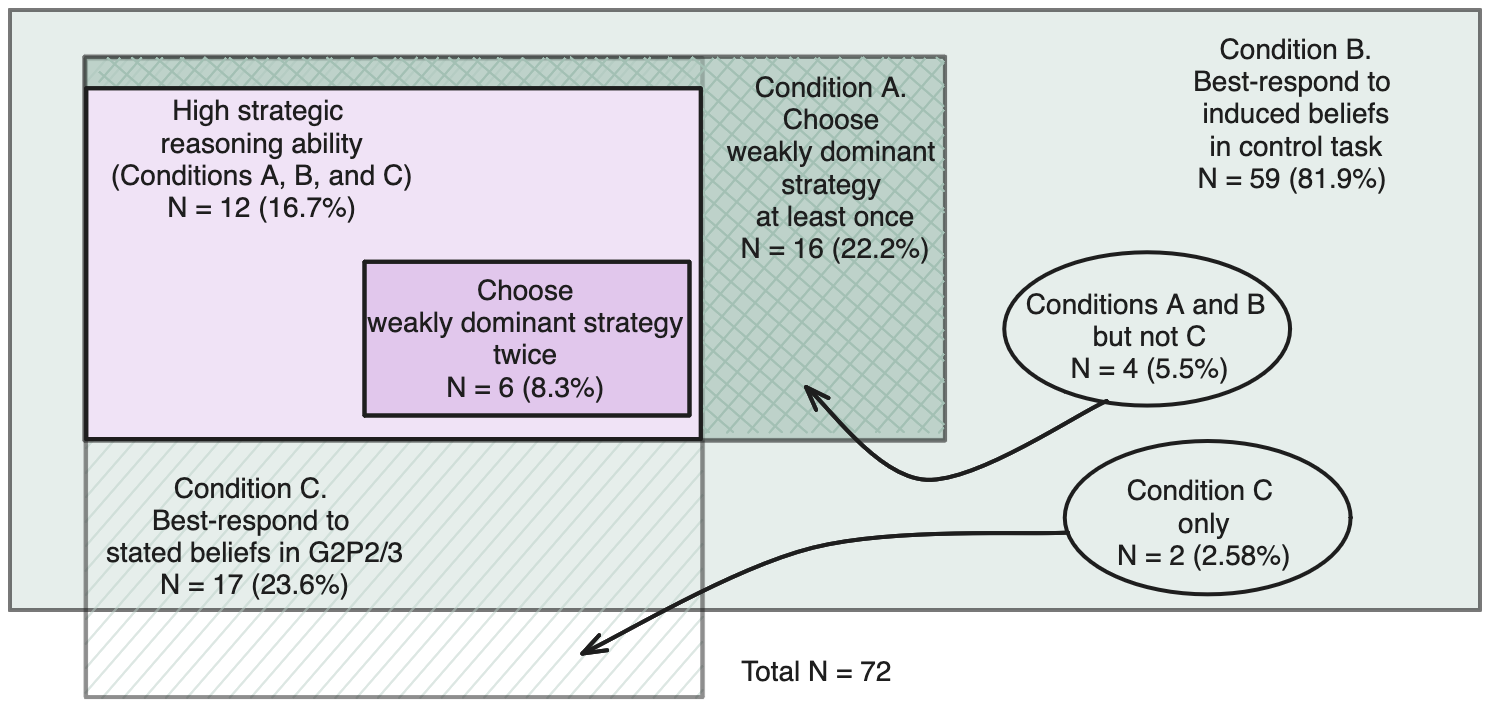}
    }
    \caption{Schema showing the number (and proportion) of subjects who meet the three conditions we impose on individual decisions to identify subjects with high strategic reasoning ability. To be classified as displaying high strategic reasoning ability, a subject must satisfy all three conditions -- A, B, and C.}
    \label{fig:high_reasoning}
\end{figure}

 \textbf{Condition B: best-respond to induced beliefs in the control task.} We believe that displaying high strategic reasoning ability requires, at a minimum, that one is able to best respond to \textit{induced} -- \textit{ie} known and given, -- beliefs. In the context of a \textit{p}-beauty contest, this entails that, if the actions of opponents are known -- meaning there is no Knightian uncertainty \cite{K_1921} or risk, -- a player with high strategic reasoning ability should be able to win the contest. 
 
 To test for ability to best-respond to induced beliefs, as explained in \Cref{sec:design_controls}, we included a control task where subjects faced a \textit{p}-beauty contests with $9$ opponents\footnote{In fact, for control, we included two versions of the contest -- with $9$ players and with $3$. Notice, however, that the latter is very similar to the understanding check that subjects had to complete after reading the instructions to participate in the experiment. We therefore consider it a weak control. Additionally, $69$ of $72$ subjects won in the $3$-player control contest, and we have no subjects winning in the $9$-player contest but failing the $3$-player one. Thus, all subject who `won' the $9$-player control contest also `won' the $3$-player one.} In this task, subjects were interacting with computerised players who were randomising on $[0,100]$ uniformly and independently. The sequence of opponents' actions was randomly generated for each subject and presented on the screen. 
 
 In our sample of $72$ subjects, $59$ ($81.9\%$) best-responded to induced beliefs in the $9$-player control contest. Moreover, of our $16$ subjects ($22.2\%$) who chose weakly dominant strategy at least once, everybody best-responded to induced beliefs, thereby satisfying the second condition.

 \textbf{Condition C: best-respond to stated beliefs in \texttt{G2P2/3}.} While being able to best-respond to induced beliefs is the minimum pre-requisite for displaying high strategic reasoning ability, to be classified as someone with such ability, one should also be capable of best-responding to their own beliefs. Given that we only have access to subjects \textit{stated} beliefs (collected through our predictions' elicitation task) rather than true beliefs, we require that the actions and predictions of subjects with high strategic reasoning ability are plausibly consistent for the \texttt{G2P2/3} task\footnote{For the $2$-player version of the contest, consistency only requires that a subjects own action is (weakly) lower than the lowest predicted action of the $17$ other experimental participants from the same session.}. We believe that best-responding to both induced and stated beliefs is unlikely to happen by accident, and so for such players stated beliefs are informative of their true beliefs.

Out of our $72$ subjects, $17$ ($23.6\%$) have plausibly consistent decisions in \texttt{G2P2/3}. This includes $5$ subjects with non-zero (but consistent) choices, and $12$ subjects with zero choices. Importantly, of the $16$ subjects who satisfy conditions A and B above, $4$ fail the requirement of best-response to stated beliefs in the \texttt{G2P2/3} version of the contest\footnote{Specifically, $3$ allocate a positive amount of tokens to bins associated with integers lower than their own chosen integer and so their decisions are clearly inconsistent. In addition, one subject allocated a positive amount of tokens to the first bin but chose $1$. It may be, that all tokens allocated to the first bin were associated with integers in $[1,5]$ rather than zero. Yet, looking at additional clarification data, we find that, if given the option, this subject would have allocated all tokens from the first bin to zero as opposed to integers in $[1,5]$, making the decision pair inconsistent. For details of the design of clarification questions see \Cref{sec:design_predictions}.}

Overall, $12$ subjects ($16.7\%$ of the sample) simultaneously satisfy all three conditions and so are classified as possessing high strategic reasoning ability.

\section{Experimental Instructions} \label{app:instructions}
Below are the instructions and understanding quizzes for all tasks from the experiment. Notice that all instructions were shown sequentially rather than together. Further, for Task \#2 (the beliefs elicitation task), participants were given paper copy of instructions in addition to those available on their computer stations. Throughout instructions and quizzes, variables that were randomised at the subject level appear in brackets -- \textit{eg} \texttt{$\langle$variable$\rangle$}. Finally, for Task \#5 (cognitive reflection test), the order of the four questions was randomised at the subject level.

\subsection*{Instructions}
Hello and welcome to our experiment!

We anticipate that completing this experiment should take about \textbf{60-90 minutes}.

Throughout the experiment, we will ask you to complete a series of \textbf{tasks} followed a short \textbf{questionnaire}. The instructions for every task will be given in due course. The questionnaire contains some questions regarding basic demographics, as well as those relating to your decisions in the experiment.

You will be paid \pounds5 for showing up to this experiment. Additionally, we will pay bonuses for your decisions in the tasks. The exact payment procedures for each task will be explained in due course.

Throughout this study you will have access to a simple on-screen \textbf{calculator}. You can access the calculator now by clicking on the `Calculator' button on the right hand-side of the interface. Should you want to make any calculations, we would like to ask you to use this calculator and \textbf{refrain from using other calculators}.

Everyone participating in this study receives the \textbf{same instructions for all tasks}.

\subsection*{Task \#1: Instructions}

Before completing Task \#1, you are asked to complete a \textbf{quiz question}. You have \textbf{3 attempts} and will only be allowed to continue with the experiment if you pass the quiz.

Once you complete the quiz question, you will be asked to make a series of \textbf{decisions in 8 decision scenarios}. We ask you to spend a minimum of \textbf{60 seconds} on each decision scenario.

For Task \#1, at the end of the experiment, we will randomly select \textbf{2 of decision scenarios} and pay you a bonus, the size of which will depend on your performance in those scenarios.

\subsection*{Task \#1: Quiz attempt \texttt{$\langle$n$\rangle$}}

Suppose you observe a group of people -- X, Y, and Z -- face the following decision scenario.

Everyone in the group is asked to choose a number between 0 and 100 (inclusive of both endpoints).

The winner is the person whose number is \textbf{closest to the target number}. The target number is calculated as \textbf{2/3rds of the average of all chosen numbers of the 3 people in the group}.

The winner gets a prize of \pounds3. In case of a tie the prize is split among those who tie.

Further suppose that X chooses \texttt{$\langle$x$\rangle$}, Y -- \texttt{$\langle$y$\rangle$}, and Z -- \texttt{$\langle$z$\rangle$}.

Who is the winner in the decision situation above?

\subsection*{Task \#2: Instructions page 1 of 3}

In the previous task, every participant was asked to make a series of decisions. In Task \#2, we would like to ask you to \textbf{estimate how the other 17 participants who are participating in this session today behaved in some of the decisions in Task \#1}.

To make your predictions, we ask you to use a set of \textbf{20 sliders}. An example of the predictions interface is below.

\begin{center}
    \includegraphics[width=0.9\textwidth]{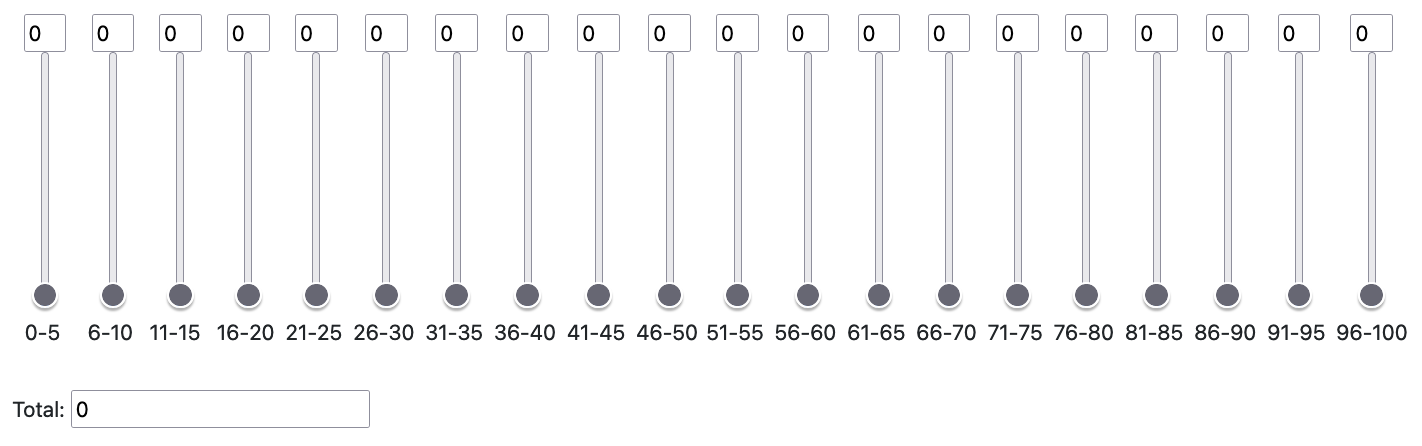}
\end{center}

As you can see, each slider represents an \textbf{interval of possible choices} -- for example, choices between 0 and 5 (inclusive) correspond to the first slider, choices between 6 and 10 -- to the second slider etc.

We ask you to allocate \textbf{17 prediction tokens} to the sliders -- one for every participant (other than you) who is participating in this session. You can allocate the tokens to sliders by either pulling the relevant sliders themselves or entering numbers directly over them using the keyboard. If you believe that no participant chose a number in a given interval, you should drag the relevant slider all the way to zero or input `0' in the field above it.

\subsection*{Task \#2: Instructions page 2 of 3}

Earlier in the experiment you faced the following decision:

\textit{Choose a number between 0 and 100 (inclusive of both endpoints).}

\textit{You will be matched in a group with 2 other anonymous participants from this session.}

\textit{The winner will be the person whose number is \textbf{closest to the target number}. The target number is calculated as \textbf{1/2 of the average of all chosen numbers of the 3 people in your group (including you).}}

\textit{The winner will get a prize of \pounds3. In case of a tie the prize will be split equally among those who tie.}

In Task \#2 you need to predict the choices that the other 17 participants from this session made in the above decision in Task \#1. As an example, suppose that you believe that in the above decision:
\begin{itemize}[noitemsep]
    \item 4 participants chose numbers between 21 and 25, and
    \item 13 participants chose numbers between 66 and 70.
\end{itemize}

In this case, you should:
\begin{itemize}[noitemsep]
    \item drag the slider labelled `21-25' to 4 (or input `4' in the field directly above that slider), and
    \item drag the slider labelled `66-70' to 13 (or input `13' in the field directly above that slider).
\end{itemize}

Feel free to use the sliders below to represent the above predictions.

\begin{center}
    \includegraphics[width=0.9\textwidth]{figures/beliefs_instructions.png}
\end{center}

\subsection*{Task \#2: Instructions page 3 of 3}

You may receive a bonus for your predictions in Task \#2. \textbf{The accuracy of your predictions determines the probability with which you receive the bonus.} In particular, once you make your predictions, we will calculate your \textbf{score} as the number of your prediction tokens that coincide with actual decisions of participants.

We will then compare your score to a \textbf{random number} drawn uniformly at random between 1 and 17 (inclusive). \textbf{If your score is equal to or greater than the random number, you win a bonus of \pounds3 for that prediction.} On the other hand, if your score is smaller than the random number, you do not win anything for that prediction. Therefore, the probability with which you win a bonus for your prediction is calculated as
       
\begin{center}
    \textbf{probability of bonus = score $\div$ 17 $\times$ 100\%}
\end{center}

As an example, recall the predictions we discussed on the previous page:
\begin{itemize}[noitemsep]
    \item 4 participants chose numbers between 21 and 25, and
    \item 13 participants chose numbers between 66 and 70.
\end{itemize}

Suppose that, indeed, 4 participants chose numbers between 21 and 25, and 13 participants chose numbers between 66 and 70. Then your score is 17. In this case, you win the bonus with probability 17 $\div$ 17 $\times$ 100\% = 100\%. In other words, you are guaranteed to win the bonus for this prediction.

On the other hand, if 13 participants chose numbers between 21 and 25, and 4 participants chose numbers between 66 and 70, then your score is 8. In this case, you win the bonus with probability 8 $\div$ 17 $\times$ 100\% = 47.1\%. In other words, your odds of winning the bonus for this prediction are just about 50:50.

Finally, if all 17 participants chose numbers between 30 and 35, then your score is 0. In this case, you win the bonus with probability 0\%. In other words, there is no chance that you win the bonus for this prediction.

In Task \#2, we will ask you to predict decisions of other participants in \textbf{3 decisions scenarios} from Task \#1. At the end of the experiment, we will randomly select \textbf{1 of your predictions} for payment. \textbf{If you won the bonus in that prediction, you would earn the value of the bonus for Task \#2.} Conversely, if you did not win a bonus for that prediction, you would earn nothing for Task \#2.

Notice that we will never pay for your choice and predictions for the same decision scenario.

Before completing Task \#2, you are asked to complete \textbf{2 quiz questions}. You have \textbf{3 attempts} and will only be allowed to continue with the experiment if you answer both questions correctly.

\subsection*{Task \#2: Quiz attempt \texttt{$\langle$n$\rangle$}}

Earlier in the experiment you faced the following decision:

\textit{Choose a number between 0 and 100 (inclusive of both endpoints).}

\textit{You will be matched in a group with 2 other anonymous participants from this session.}

\textit{The winner will be the person whose number is \textbf{closest to the target number}. The target number is calculated as \textbf{1/2 of the average of all chosen numbers of the 3 people in your group (including you)}.}

\textit{The winner will get a prize of \pounds3. In case of a tie the prize will be split equally among those who tie.}

\textbf{Question A.} Suppose that you believe that in the above decision:
\begin{itemize}[noitemsep]
    \item \texttt{$\langle$prediction 1$\rangle$} participants chose numbers between \texttt{$\langle$range 1$\rangle$},
    \item \texttt{$\langle$prediction 2$\rangle$} participants chose numbers between \texttt{$\langle$range 2$\rangle$}, and
    \item \texttt{$\langle$prediction 3$\rangle$} participants chose numbers between \texttt{$\langle$range 3$\rangle$}.
\end{itemize}

Further suppose that the other participants choices were actually as follows:
\begin{itemize}[noitemsep]
    \item \texttt{$\langle$actual 1$\rangle$} participants chose numbers between \texttt{$\langle$range 1$\rangle$},
    \item \texttt{$\langle$actual 2$\rangle$} participants chose numbers between \texttt{$\langle$range 2$\rangle$}, and
    \item \texttt{$\langle$actual 3$\rangle$} participants chose numbers between \texttt{$\langle$range 3$\rangle$}.
\end{itemize}

Please calculate your score based on your predictions and actual participants choices.

\textbf{Question B.} Suppose your score is \texttt{$\langle$score$\rangle$} and the random number is \texttt{$\langle$x$\rangle$}. Do you win the bonus for your predictions?

\subsection*{Task \#3: Instructions}

Task \#3 contains \textbf{1 question} similar to the decision situations you have encountered earlier in the experiment. For this task, we will pay you a bonus the size of which will depend on your decision in the question. You will have access to the calculator in the usual part of the interface.

\subsection*{Task \#3: Question}

Suppose you face the following decision scenario. You need to choose a number between 0 and 100 (inclusive of both endpoints). You will be matched in a group with 2 other anonymous players.

The winner will be the player whose number is closest to the target number. The target number is 2/3rds of the average of all chosen numbers of the 3 players in your group (including you). The winner will get a prize of \pounds3. In case of a tie the prize will be split equally among those who tie.

Further suppose that the 2 anonymous players you are interacting with are advanced language models. In particular, you are playing against two advanced generative artificial intelligence chatbots: Chat-GPT (developed by OpenAI) and Claude (developed by Anthropic).

Choose your number.

\subsection*{Task \#4: Instructions}

Task \#4 contains \textbf{2 questions} similar to the decision scenarios you have encountered earlier in the experiment. We will \textbf{randomly select 1 question and pay you a bonus}, the size of which will depend on your decision in that question. You will have access to the calculator in the usual part of the interface.

\subsection*{Task \#4: Questions}

\textbf{Question 1.} Suppose you face the following decision scenario. You need to choose a number between 0 and 100 (inclusive of both endpoints). You will be matched in a group with 2 other anonymous players.

The winner will be the player whose number is closest to the target number. The target number is 2/3rds of the average of all chosen numbers of the 3 players in your group (including you). The winner will get a prize of \pounds3. In case of a tie the prize will be split equally among those who tie.

Further suppose that the choices of other players are known to you. Specifically, they choose the following numbers: \texttt{$\langle$sequence of 2 random integers on $[0,100]$$\rangle$}. Choose your number.

\textbf{Question 2.} Suppose you face the following decision scenario. You need to choose a number between 0 and 100 (inclusive of both endpoints). You will be matched in a group with 8 other anonymous players.

The winner will be the player whose number is closest to the target number. The target number is 2/3rds of the average of all chosen numbers of the 9 players in your group (including you). The winner will get a prize of \pounds3. In case of a tie the prize will be split equally among those who tie.

Further suppose that the choices of other players are known to you. Specifically, they choose the following numbers: \texttt{$\langle$sequence of 8 random integers on $[0,100]$$\rangle$}. Choose your number.

\subsection*{Task \#5: Instructions}

Task \#5 consists of \textbf{4 questions}. You will earn \textbf{\pounds1 for each correct answer} and nothing for each incorrect answer. You have 4 minutes to complete this task and the timer will always be displayed at the top of the page. You will have access to the calculator in the usual part of the interface.

\textbf{Task \#5: Question 1 of 4. } If you’re running a race and you pass the person in 2nd place, what place are you in?

\textbf{Task \#5: Question 2 of 4.} A farmer had 15 sheep and all but 8 died. How many are left?

\textbf{Task \#5: Question 3 of 4.} Jerry received both the 15th highest and the 15th lowest mark in the class. How many students are in the class?
        
\textbf{Task \#5: Question 4 of 4.} A man buys a pig for \pounds60, sells it for \pounds70, buys it back for \pounds80, and sells it finally for \pounds90. How much has he made?

\section{Screenshots from the interface} \label{app:screenshots}
\begin{figure}[h!]
    \centering
    \frame{
        \includegraphics[width=0.4\textwidth]{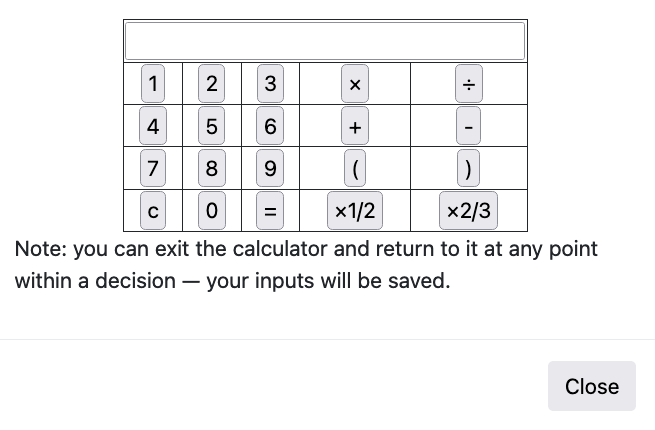}
    }
    \caption{Interface of the on-screen calculator used in the experiment.}
    \label{fig:calculator_interface}
\end{figure}

\begin{figure}[h!]
    \centering
    \frame{
        \includegraphics[width=0.9\textwidth]{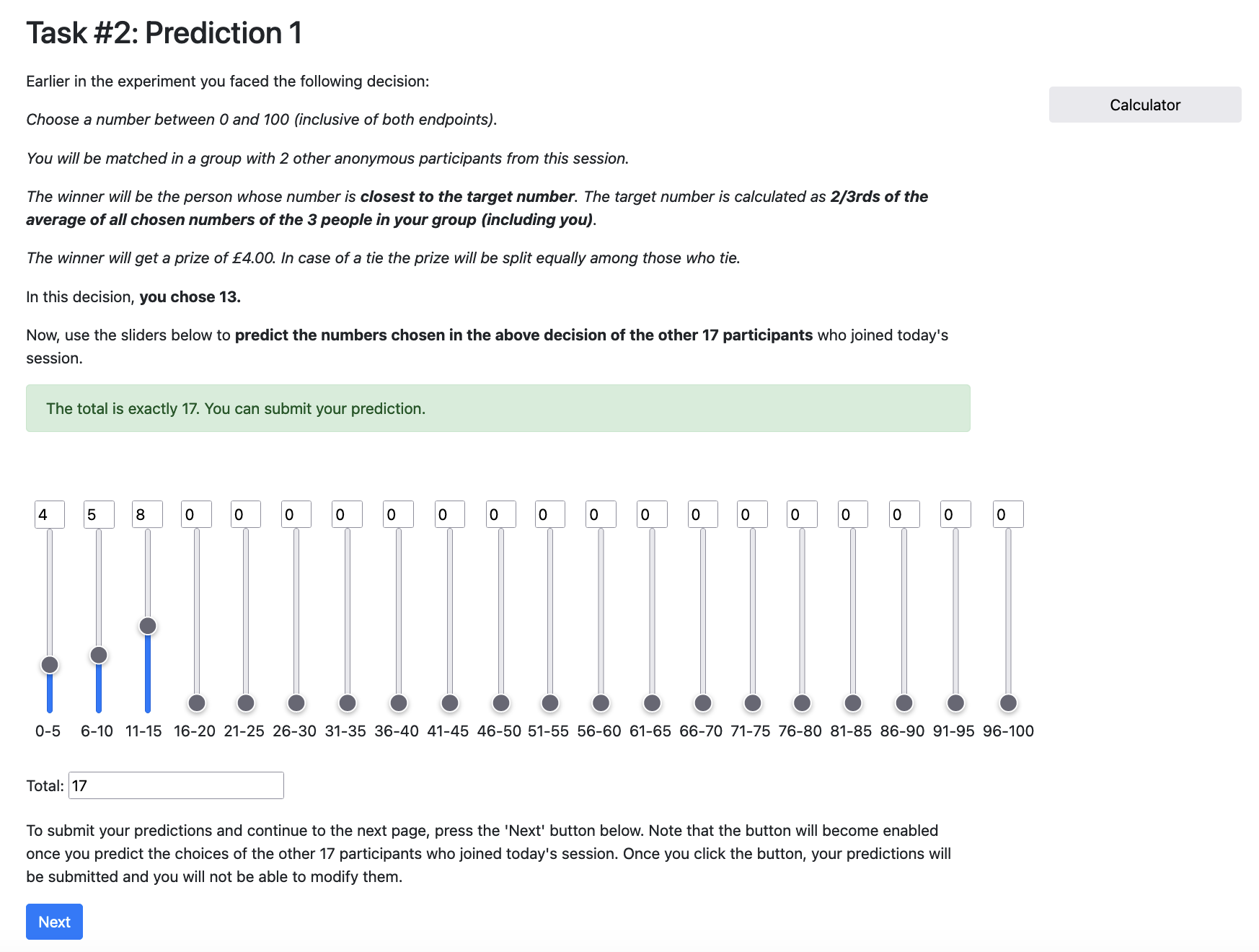}
    }
    \caption{Interface of the predictions' elicitation task of the experiment.}
    \label{fig:predictions_interface}
\end{figure}

\end{document}